\documentclass[a4paper, amsfonts, amssymb, amsmath, reprint, showkeys, nofootinbib, superscriptaddress]{revtex4-1}
\usepackage[english]{babel}
\usepackage[utf8]{inputenc}
\usepackage[colorinlistoftodos, color=green!40, prependcaption]{todonotes}
\usepackage{amsthm}
\usepackage{mathtools}
\usepackage{physics}
\usepackage{xcolor}
\usepackage{graphicx}
\usepackage[left=23mm,right=13mm,top=35mm,columnsep=15pt]{geometry} 
\usepackage{adjustbox}
\usepackage{placeins}
\usepackage[T1]{fontenc}
\usepackage{lipsum}
\usepackage{csquotes}
\usepackage[pdftex, pdftitle={Article}, pdfauthor={Author}]{hyperref} 
\begin{document}
\title{Quantifying the efficiency of principal signal transmission modes in proteins}

\author{Anil Kumar Sahoo}
    \affiliation{Fachbereich Physik, Freie Universit\"{a}t Berlin, Arnimallee 14, 14195 Berlin, Germany}
    \affiliation{Max Planck Institute of Colloids and Interfaces, Am M\"{u}hlenberg 1, 14476 Potsdam, Germany}

\author{Richard Schwarzl}
    \affiliation{Fachbereich Physik, Freie Universit\"{a}t Berlin, Arnimallee 14, 14195 Berlin, Germany}

\author{Markus S. Miettinen}
    \altaffiliation[Present address: ]{ Computational Biology Unit, Departments of Chemistry and Informatics, University of Bergen, 5007 Bergen, Norway}
    \affiliation{Fachbereich Physik, Freie Universit\"{a}t Berlin, Arnimallee 14, 14195 Berlin, Germany}

\author{Roland R. Netz}
    \email[]{rnetz@physik.fu-berlin.de}
    \affiliation{Fachbereich Physik, Freie Universit\"{a}t Berlin, Arnimallee 14, 14195 Berlin, Germany}
    \affiliation{Centre for Condensed Matter Theory, Department of Physics, Indian Institute of Science, Bangalore 560012, India}


\begin{abstract} 
On the microscopic level,
biological signal transmission relies on coordinated structural changes in  allosteric proteins that involve  sensor and effector modules. 
 The timescales and microscopic  details of  signal transmission in proteins are often unclear, despite a plethora of structural information on signaling proteins. 
 Based on linear-response theory, we develop a theoretical framework  to define frequency-dependent force and displacement transmit functions
 through proteins and, more generally,  viscoelastic media. 
Transmit functions quantify the fraction of a local time-dependent perturbation at one site, be it a deformation,
a force or a combination thereof, that survives at a second site. They are defined in terms
of equilibrium fluctuations from simulations or experimental observations.
 We apply the framework to our all-atom molecular dynamics simulation data of a parallel, homodimeric coiled-coil (CC) motif that connects the sensor and effector modules of a
 blue-light-regulated histidine kinase from bacterial signaling systems extensively studied in experiments. 
 Our analysis reveals that signal transmission through the CC is possible via shift, splay, and twist deformation modes. 
Based on results of mutation experiments, we infer that the most relevant mode for the biological function of the histidine kinase protein is the splay deformation.
\end{abstract}


\maketitle

\section*{Introduction}
Signal transmission within a cell, between cells, and from the exterior to the interior of a cell is necessary for any form of life, be it bacteria, plants, or animals. Signal-transducing units, at the molecular level, involve signal receptors (sensors) that sense intracellular or environmental changes (such as photons, hormones or pH) and corresponding effectors that are responsible for sparking a response \cite{hunter2000signaling, smock2009sending}.
Sensor and effector modules are usually distinct domains of membrane-bound or cytosolic proteins \cite{rosenbaum2009structure, bhate2015signal, gleichmann2013charting}. How information, at the molecular level, is transmitted through  proteins has been subject of intense research in the last few decades \cite{wodak2019allostery, cooper1984allostery, suel2003evolutionarily, changeux2005allosteric, cui2008allostery, motlagh2014ensemble, nussinov2014principles, feher2014computational, thirumalai2019symmetry}. Experimental techniques, such as NMR spectroscopy \cite{bruschweiler2009direct, guo2016protein}, time-resolved crystallography \cite{berntsson2017sequential, mehrabi2019time}, cryo-electron microscopy \cite{wahlgren2022structural}, and single-molecule experiments \cite{lerner2018toward, schoeler2015mapping, bauer2019structural} have provided information about protein structure, dynamics, and mechanical signaling pathways. Computational approaches combining molecular simulation techniques \cite{miyashita2003nonlinear, zheng2006low, van2006interpreting, sharp2006pump, young2013fast} and tools from information theory \cite{lenaerts2008quantifying} and graph theory \cite{chennubhotla2006markov, sethi2009dynamical, negre2018eigenvector}, along with various linear or non-linear correlation analyses \cite{lange2006generalized, kong2009signaling}, have provided molecular-level insights into protein allosteric communication pathways \cite{mitchell2016strain, wang2020mapping}.  A combination of spectroscopy techniques and molecular dynamics (MD) simulation description has led to recent insights into time-resolved processes of allosteric regulation \cite{buchli2013kinetic, viappiani2014experimental, buchenberg2017time, bozovic2020real, bozovic2021speed, ali2022nonequilibrium}. 
However, the quantitative  relation between  the dynamics of a protein and its signal transfer efficacy, which ultimately governs physiological response, is missing. \par{}
We here show how to  quantify the  signal-transfer efficiency through proteins in terms of  frequency-dependent force and  displacement  transmit functions. A force  transmit function describes the fraction of  the frequency-dependent  force applied at the protein sensor position that survives at the effector position. The displacement transmit function is defined similarly but is based on spatial displacements.  
In fact, the presence of strongly correlated fluctuations  at the sensor and effector positions, as
quantified by two-point correlation functions,  is not sufficient for efficient  signal transmission, 
because equilibrium fluctuations produce a background that the signal has to compete with. 
Transmit functions weigh the correlation between the sensor and effector positions by the fluctuation
magnitude and therefore quantify the fractional signal transmission, they are thus very different from ordinary two-point correlation functions.
Our theoretical framework builds on our previously developed convolution theory 
 \cite{hinczewski2010deconvolution} and is exact on the linear-response level \cite{kubo1966fluctuation}. 
 From the time-domain transmit function, the protein response to any temporal perturbation signal can
 be calculated by convolution, all one needs as input to our theory are time series of positions or
 displacements that can be obtained from MD simulations or single-molecule experiments,
  e.g., by fluorescence resonance energy transfer \cite{lerner2018toward}.\par{} 
Here, we focus on coiled-coil motifs, frequently found in various plant and bacterial
signal transduction systems, that connect signal receptor and response or effector protein \cite{lupas2005structure, casino2010mechanism, anantharaman2006signaling, matthews2006dynamic, hulko2006hamp, inouye2006signaling, bhate2015signal, moglich2019signal}. Signal receptors are commonly oligomeric domains and modular in architecture, produced by recombination of sensor and effector modules. Signaling through coiled coils leads to conformational rearrangements within effectors that trigger interaction with regulator proteins, starting signaling cascades.
Studies investigating the role of coiled coils in signal propagation are, however, limited to quantifying their structural changes in the presence of an external signal, e.g., pivoting or rotation of the helices \cite{wahlgren2022structural, moglich2019signal}.\par
\begin{figure}[t]
    \centering   \includegraphics[width=0.5\textwidth]{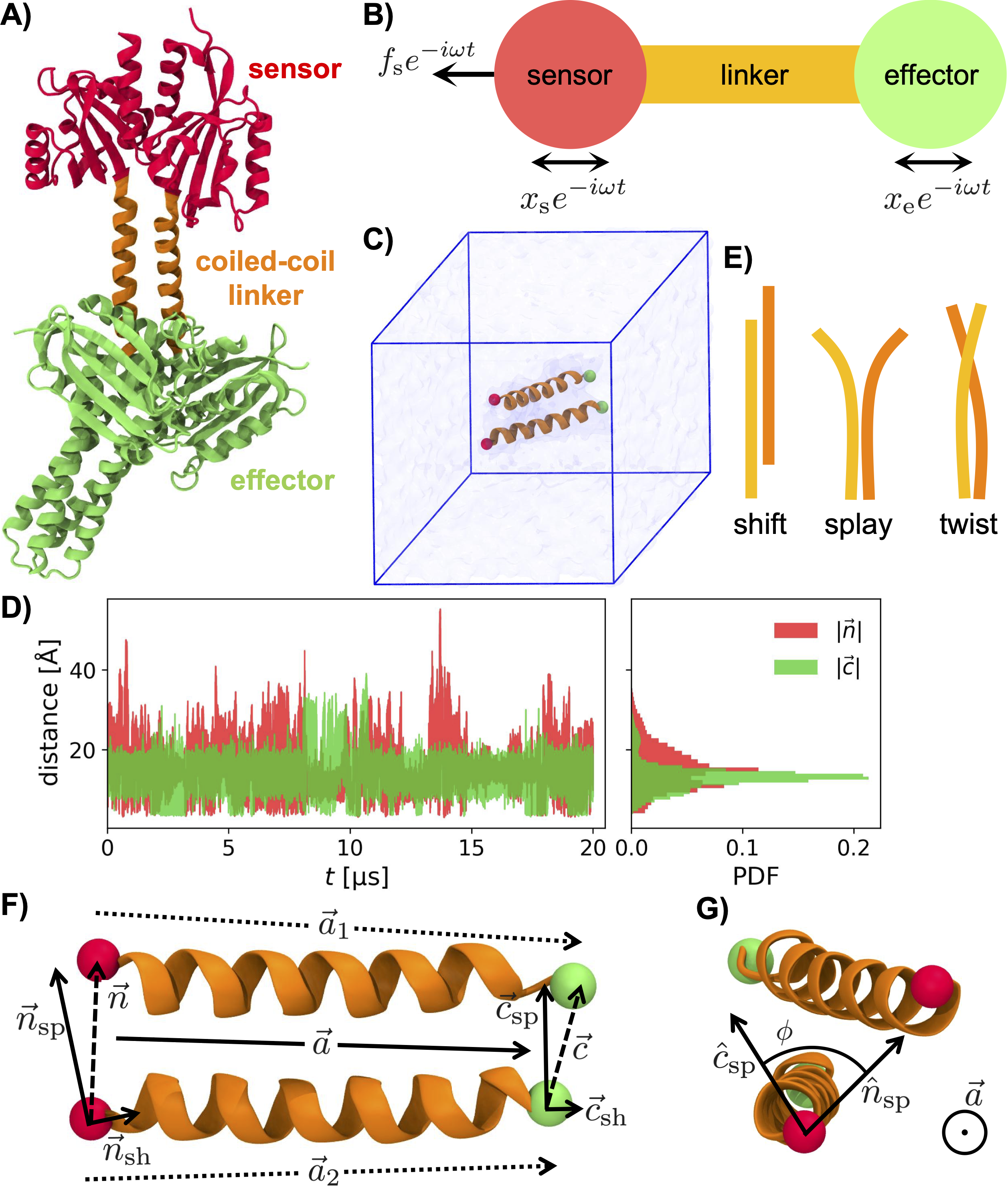}
    \caption{\label{Fig1} A) The full-length structure of the dark-adapted blue-light-regulated  histidine kinase YF1 (PDB ID: 4GCZ) \cite{diensthuber2013full}. In YF1 the sensor, linker, and effector modules are homodimeric at the molecular level.
    B) Schematic representation of  signal transmission from sensor to effector sites. C) Simulation unit cell containing the coiled-coil (CC) linker. 
    Water and ions are not shown. D) Time series and corresponding probability density functions (PDFs) for the N and C-termini distances of the CC, 
   $|\vec{n}|$ and  $|\vec{c}|$. E) Schematic representation of shift, splay, and twist modes of the CC. F) Schematics for defining the shift vectors $\vec{n}_\mathrm{sh}$, $\vec{c}_\mathrm{sh}$
    and the splay vectors $\vec{n}_\mathrm{sp}$,  $\vec{c}_\mathrm{sp}$ as the parallel and perpendicular components of the distance vectors $\vec{n}$, $\vec{c}$ with respect to the symmetrized CC axis $\vec{a} = (\vec{a}_\mathrm{1} + \vec{a}_\mathrm{2})/2$. G) Schematic for describing twist as rotation of splay unit vectors $\hat{n}_\mathrm{sp}$  and 
    $\hat{ c}_\mathrm{sp}$ around the long-axis $\vec{a}$. The angle between $\hat{n}_\mathrm{sp}$  and $\hat{ c}_\mathrm{sp}$ is denoted as the twist angle $\phi$.
    }
    \end{figure}

{}
We perform all-atom MD simulations  of a coiled-coil (CC) motif consisting of two identical $\alpha$-helices arranged in parallel that
is part of an engineered  blue-light-regulated histidine kinase \cite{moglich2009design}. 
In this synthetic enzyme, 
the CC connects the light-oxygen-voltage sensor module from Bacillus subtilis YtvA and the histidine kinase complex from the bacterium Bradyrhizobium japonicum,
see Fig. \ref{Fig1}A.
We study the dynamics of the isolated  CC in terms of its shift, splay, and twist deformation modes. 
Computed transmit functions reveal that all modes are able to transmit signals. 
From the time-domain transmit functions we derive transmit properties of different time-dependent signals. 
For step signals we find that transmission via the splay mode is drastically reduced by single-point
mutations in the CC, whereas the twist mode is least affected. Together with the experimental
observation of light-induced splaying of the CC \cite{berntsson2017sequential, berntsson2017time} 
and experimental mutation studies \cite{gleichmann2013charting}, this suggests that splay is 
the most relevant deformation mode of the CC for the biological function of this sensor histidine kinase \cite{moglich2009design}.\par{}
%
\section*{Results}
\subsection{Theory of transmit functions}
    The input--output relation for a general responsive system can be quantified by two transmit functions: force and displacement transmission.
    Consider we apply to a system's sensor (input) position a frequency-dependent force $\Tilde{F}_\mathrm{s}(\omega)$ 
    and to its effector (output) position  a force $\Tilde{F}_\mathrm{e}(\omega)$, see Fig. \ref{Fig1}B.
    In Fourier space, the sensor and effector positions change as \cite{hinczewski2010deconvolution}
    \begin{eqnarray}\label{Eq1}
    \begin{split}
        \Tilde{X}_\mathrm{s}(\omega) = \Tilde{J}^\mathrm{s}_\mathrm{self}(\omega)\Tilde{F}_\mathrm{s}(\omega) + \Tilde{J}_\mathrm{cross}(\omega)\Tilde{F}_\mathrm{e}(\omega) \\
        \Tilde{X}_\mathrm{e}(\omega) = \Tilde{J}^\mathrm{e}_\mathrm{self}(\omega)\Tilde{F}_\mathrm{e}(\omega) + \Tilde{J}_\mathrm{cross}(\omega)\Tilde{F}_\mathrm{s}(\omega),
    \end{split}
    \end{eqnarray}

    where $\Tilde{J}_\mathrm{cross} (\omega)$ and $\Tilde{J}^\mathrm{s/e}_\mathrm{self} (\omega)$ are the frequency-dependent cross and self (sensor/effector-side) linear response functions, respectively. Note that there is only a single cross-response. 
However, the self-responses $\Tilde{J}^\mathrm{s}_\mathrm{self}(\omega)$ and $\Tilde{J}^\mathrm{e}_\mathrm{self}(\omega)$ are different for a  general asymmetric system.  
It should be further noted that Eq. \ref{Eq1} is exact on the linear response level and is applicable for any two positions in a protein.    
The force transmit function $\Tilde{T}^\mathrm{s \rightarrow e}_\mathrm{F}(\omega)$ is defined for the boundary condition of a stationary effector position
 $\Tilde{X}_\mathrm{e} = 0$, from Eq. \ref{Eq1}, we obtain 
    \begin{eqnarray}\label{Eq3}
   \Tilde{T}^{\mathrm{s \rightarrow e}}_\mathrm{F} (\omega) \equiv \frac{\Tilde{F}_\mathrm{e} (\omega)}{-\Tilde{F}_\mathrm{s} (\omega)} = \frac{\Tilde{J}_\mathrm{cross}(\omega)}{\Tilde{J}^\mathrm{e}_\mathrm{self}(\omega)}.
    \end{eqnarray}
    Similarly, setting $\Tilde{X}_\mathrm{s} = 0$ in Eq. \ref{Eq1}, we obtain $\Tilde{T}^\mathrm{e \rightarrow s}_\mathrm{F} 
    \equiv      - \Tilde{F}_\mathrm{s} /  \Tilde{F}_\mathrm{e} = \Tilde{J}_\mathrm{cross}/\Tilde{J}^\mathrm{s}_\mathrm{self}$.
    Response functions are \textit{causal}, i.e., there is no positional response before a force is applied, 
    which implies $\Tilde{J}_\mathrm{cross}(\omega)$ and $\Tilde{J}_\mathrm{self}(\omega)$ are analytic in the upper-half complex plane. If furthermore
    $\Tilde{J}_\mathrm{self}(\omega)$ is non-zero in the upper-half complex plane, from Eq. \ref{Eq3} it follows that transmit functions are  also causal.\par{}
    The displacement transmit function  $\Tilde{T}^\mathrm{s \rightarrow e}_\mathrm{X}(\omega)$ 
    is defined as  the ratio of  the displacement of the  effector  site $\Tilde{X}_\mathrm{e}$ divided by 
    the  displacement at the sensor  site  $\Tilde{X}_\mathrm{s}$ under force-free  boundary condition $\Tilde{F}_\mathrm{e} = 0$. 
    Inverting Eq. \ref{Eq1} yields 
    \begin{eqnarray}\label{Eq4}
      \begin{split}
        \begin{pmatrix} \Tilde{F}_\mathrm{s} \\ \Tilde{F}_\mathrm{e} \end{pmatrix}
        & =
        \frac{1}{\Tilde{J}^\mathrm{s}_\mathrm{self}\Tilde{J}^\mathrm{e}_\mathrm{self}-{\Tilde{J}_\mathrm{cross}}^2}
        \begin{pmatrix} \Tilde{J}^\mathrm{e}_\mathrm{self} &  -\Tilde{J}_\mathrm{cross} \\ -\Tilde{J}_\mathrm{cross} & \Tilde{J}^\mathrm{s}_\mathrm{self} \end{pmatrix}
        \begin{pmatrix} \Tilde{X}_\mathrm{s} \\ \Tilde{X}_\mathrm{e} \end{pmatrix} \\
         & =
        \begin{pmatrix} \Tilde{G}^\mathrm{s}_\mathrm{self} &  \Tilde{G}_\mathrm{cross} \\ \Tilde{G}_\mathrm{cross} & \Tilde{G}^\mathrm{e}_\mathrm{self}  \end{pmatrix}        
        \begin{pmatrix} \Tilde{X}_\mathrm{s} \\ \Tilde{X}_\mathrm{e} \end{pmatrix}, 
      \end{split} 
    \end{eqnarray}  
    where $\Tilde{G}$'s are the  moduli determined by  inverting the response matrix 
    $\begin{pmatrix} \Tilde{J}^\mathrm{s}_\mathrm{self} &  \Tilde{J}_\mathrm{cross} \\ \Tilde{J}_\mathrm{cross} & \Tilde{J}^\mathrm{e}_\mathrm{self} \end{pmatrix}$.
   From  Eq. \ref{Eq4} we obtain 
    \begin{eqnarray}\label{Eq5}
    \Tilde{T}^\mathrm{s \rightarrow e}_\mathrm{X} \equiv 
         \frac{\Tilde{X}_\mathrm{e}}{\Tilde{X}_\mathrm{s}} = \frac{-\Tilde{G}_\mathrm{cross}}{\Tilde{G}^\mathrm{e}_\mathrm{self}} = \frac{\Tilde{J}_\mathrm{cross}}{\Tilde{J}^\mathrm{s}_\mathrm{self}} = \Tilde{T}^\mathrm{e \rightarrow s}_\mathrm{F}      
    \end{eqnarray}
 and by setting  $\Tilde{F}_\mathrm{s} = 0$ similarly  $ \Tilde{T}^\mathrm{e \rightarrow s}_\mathrm{X} \equiv   \Tilde{X}_\mathrm{s} / \Tilde{X}_\mathrm{e} =
     \Tilde{T}^\mathrm{s \rightarrow e}_\mathrm{F}$.     
    Thus, force and inverse displacement transmit functions are the same. 
 Note the striking resemblance between   transmit functions defined here  and the transfer function 
 which characterizes the output of a linear time-invariant system (e.g., an electric circuit consisting of resistors, inductors and capacitors) in the context of signal processing \cite{oppenheim1996signals}. \par{}
    In practice, one need not apply external forces to determine linear response functions. $J(t)$ can be obtained from equilibrium
     time-correlation functions $C(t)$  using the fluctuation--dissipation theorem \cite{kubo1966fluctuation}:
    \begin{eqnarray}\label{Eq6}
        J(t) = -\frac{1}{k_\mathrm{B}T}\theta(t)\frac{d}{dt}C(t),
    \end{eqnarray}
    where $k_\mathrm{B}$ is the Boltzmann's constant, $T$ represents temperature, and $\theta(t)$ is the Heaviside step function. 
   The needed cross and  self correlation functions are defined as $C_\mathrm{cross}(t) = \langle X_\mathrm{s}(0)X_\mathrm{e}(t)\rangle$ and
   $C_\mathrm{self}^\mathrm{s}(t) = \langle X_\mathrm{s}(0)X_\mathrm{s}(t)\rangle$, $C_\mathrm{self}^\mathrm{e}(t) = \langle X_\mathrm{e}(0)X_\mathrm{e}(t)\rangle$, respectively (see  \textit{Materials and Methods} for details).
    Note that the  trajectories of the  positions of two sites in a protein needed for the calculation of $C(t)$ can be obtained from MD simulations but also from 
   single-molecule experiments \cite{lerner2018toward}. \par{} 
\subsection{MD simulations of the CC linker}
    We consider the  bacterial signaling protein introduced  above to demonstrate the applicability of our theoretical framework to real systems. 
   The water-soluble structure of the whole protein (Fig. \ref{Fig1}A) is too large to carry out simulations for a sufficiently long time. 
   Also, we are primarily interested in the CC because of its relevance in diverse signal receptors \cite{lupas2005structure, anantharaman2006signaling, bhate2015signal, moglich2019signal} and of the availability of structural and experimental data \cite{moglich2009design, diensthuber2013full, gleichmann2013charting, berntsson2017sequential, berntsson2017time, ohlendorf2016library}. 
   We therefore perform explicit solvent all-atom MD simulations of only the CC linker excluding the sensor and effector modules.
    A summary of simulation details is given in \textit{Materials and Methods} and the simulation unit cell is shown in Fig. \ref{Fig1}C. 
    We find that the distribution of  the N-termini  distance at the sensor side is  broader compared to the C-termini  distance  at the effector side,
    both distributions exhibit tails that reflect
   intermittent splaying of the $\alpha$-helix termini  (see Fig. \ref{Fig1}D,E).
To check the long-time stability of secondary and tertiary structures of the CC, we calculate three different order parameters: 
the fraction of native contacts $Q$ between the two $\alpha$-helices, the root-mean-square deviation (RMSD) of distances  between
the native and simulated CC structures, the former taken from the crystal structure of the full-length protein shown in Fig. \ref{Fig1}A, 
and the secondary structure (SS) content. These order parameters are defined in \textit{SI Appendix}, section 1
and their time-averaged values are given in \textit{SI Appendix}, Fig. S1. 
The overall configuration of the CC remains stable within 20 $\mu$s of simulation as indicated by an average RMSD of 2.5 \AA{}. The two $\alpha$-helices remain bound to each other ($Q = 0.92$) due to salt bridges (involving residues Arg$^{135}$ and Glu$^{138, 142}$) and hydrophobic interactions (involving residues Leu$^{136, 139, 143}$ and Val$^{146}$) \cite{gleichmann2013charting, diensthuber2013full}. Individual $\alpha$-helices do not melt as their fraction of SS content values is above 0.85. \par{} 
\subsection{The CC linker transmits signals via shift, splay and twist modes}
 We consider the CC linker deformations that correspond to forces of equal magnitude and opposite
 direction acting on the two N-termini and on the two C-termini. Thus, these deformations conserve
 linear momentum. We define three distinct deformation modes by the orientation of the terminal
 displacement vector with respect to the distance vector between the terminal groups:
 splay, where the displacement is parallel to the terminal separation, and shift and twist, where the 
 displacement is perpendicular to the terminal separation, see schematics in Fig. \ref{Fig1}E.
 Splay conserves angular momentum, whereas shift and twist do not and are counteracted by a rotation of the sensor module. 
 Since the typical rotational diffusion time of the sensor module, estimated to be of the order of 
 1 $\mu$s (see \textit{Materials and Methods}), is much longer than shift and twist relaxation times,
 as will be shown below, shift and twist modes are nevertheless possible signal transmission modes. \par{}
    \begin{figure*}[t]
    \centering
    \includegraphics[width=0.74\textwidth]{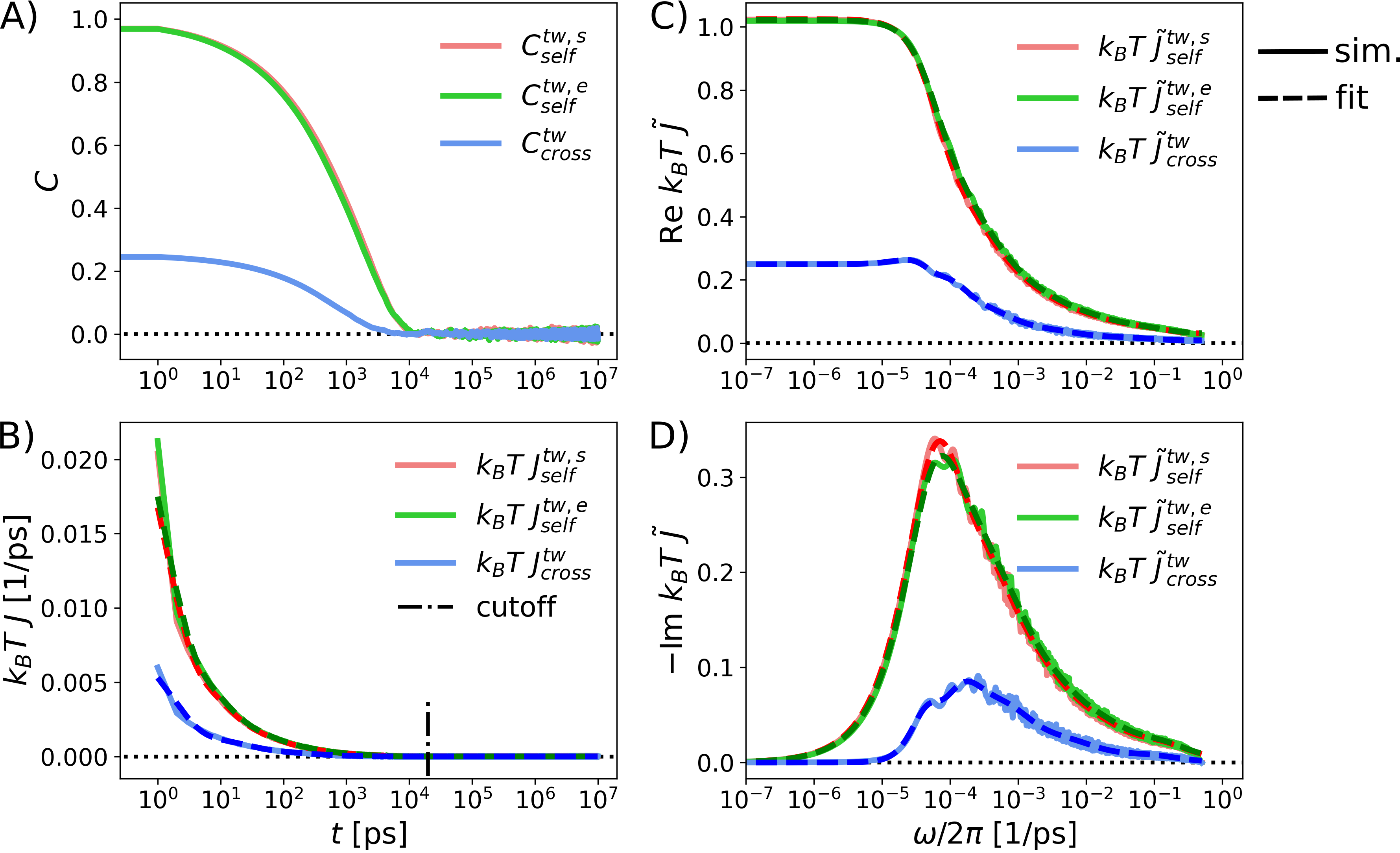}
    \caption{\label{Fig2} A) Self and cross-correlation functions ($C_\mathrm{self}$ and $C_\mathrm{cross}$) for the twist mode of the CC. 
    B)  Response functions $J$, obtained according to  Eq. \ref{Eq6} from numerical  derivatives of the correlation functions in A  (solid lines). 
    The vertical dash-dotted line represents the cutoff beyond which the response functions are set to zero to prevent noise artifacts when calculating Fourier transforms.
   C) Real and D) imaginary parts of $\Tilde{J}(\omega)$ obtained from discrete Fourier transform of  $J(t)$ are shown as solid lines. 
   Dashed lines represent simultaneous  fits to the real and imaginary parts  by a sum of 10 Debye relaxation functions (for further details see \textit{SI Appendix}, section 3
    ). Dashed lines in panel B represent the inverse Fourier transform of the fits in C and D.} 
    \end{figure*}
  The three signal transmission modes are obtained as described below.
  The position vectors of the  N-termini and C-termini are $\vec{N}_\mathrm{i}$ and $\vec{C}_\mathrm{i}$, respectively, where $i = 1, 2$ refers to the first and second $\alpha$-helix. These vectors are used to construct the separation vectors $\vec{c} = \vec{C}_\mathrm{2} - \vec{C}_\mathrm{1}$, $\vec{n} = \vec{N}_\mathrm{2} - \vec{N}_\mathrm{1}$, and $\vec{a}_\mathrm{i} = \vec{C}_\mathrm{i} - \vec{N}_\mathrm{i}$ (Fig. \ref{Fig1}F). From the end-to-end vectors $\vec{a}_\mathrm{i}$, we define the long axis of the CC as $\vec{a} = (\vec{a}_\mathrm{1} + \vec{a}_\mathrm{2})/2$. With respect to  $\vec{a}$, we separate $\vec{n}$ into the parallel component, shift ($n_\mathrm{sh} = |\vec{n}_\mathrm{sh}| = \vec{n} \cdot \hat{a}$) and  the perpendicular component, splay ($n_\mathrm{sp} = |\vec{n}_\mathrm{sp}| = |\vec{n}-\vec{n}_\mathrm{sh}|$). 
    Similarly, for the C-termini we obtain $c_\mathrm{sh}$ and $c_\mathrm{sp}$. The twist angle 
    $\phi$, depicted in Fig. \ref{Fig1}G, is obtained from the scalar product of the two splay
    unit vectors $\hat{n}_\mathrm{sp}$ and $\hat{c}_\mathrm{sp}$ as $\phi = \cos^{-1}(\hat{n}_\mathrm{sp} \cdot \hat{c}_\mathrm{sp})$. We show time series of these deformation modes and the corresponding probability distribution functions in \textit{SI Appendix}, Fig. S2. 
  Importantly, we find a strong bias towards clockwise rotation of the CC around its long axis $\vec{a}$, from the handedness plot obtained from the simulation (see \textit{SI Appendix}, Fig. S2), in agreement with the experimental observation of the left-handed supercoiling of the CC \cite{berntsson2017sequential}. \par{}  
    The N and C-termini  of the CC correspond to the sensor (s) and effector (e) side, respectively.
    For the shift mode, the two self-correlation functions are defined as $C_\mathrm{self}^\mathrm{sh, s} = \langle n_\mathrm{sh}(0)n_\mathrm{sh}(t)\rangle$ and $C_\mathrm{self}^\mathrm{sh, e} = \langle c_\mathrm{sh}(0)c_\mathrm{sh}(t)\rangle$ and the cross-correlation function is defined as 
    $C_\mathrm{cross}^\mathrm{sh} = \langle n_\mathrm{sh}(0)c_\mathrm{sh}(t)\rangle$     $= \langle c_\mathrm{sh}(0)n_\mathrm{sh}(t)\rangle$.     
    Correlation functions for the splay mode are defined by interchanging the shift quantities with the related splay quantities, e.g., $C_\mathrm{self}^\mathrm{sp, s} = \langle n_\mathrm{sp}(0)n_\mathrm{sp}(t)\rangle$. For the twist mode, the cross and two self-correlation functions are defined by the scalar product of the splay unit vectors 
    as $C_\mathrm{cross}^\mathrm{tw} = \langle \hat{n}_\mathrm{sp}(0) \cdot \hat{c}_\mathrm{sp}(t)\rangle = \langle \hat{c}_\mathrm{sp}(0) \cdot \hat{n}_\mathrm{sp}(t)\rangle$ and $C_\mathrm{self}^\mathrm{tw, s} = \langle \hat{n}_\mathrm{sp}(0) \cdot \hat{n}_\mathrm{sp}(t)\rangle$ and $C_\mathrm{self}^\mathrm{tw, e} =
     \langle \hat{c}_\mathrm{sp}(0) \cdot \hat{c}_\mathrm{sp}(t)\rangle$, respectively. 
    To disentangle twist from overall CC rotation, we calculate  twist correlation functions in the
    molecular coordinate frame obtained by removing the CC center-of-mass translation and
    rigid-body rotation around its principal axes at each time step.
    We present results for the twist mode in Fig. \ref{Fig2} and for shift and splay modes in \textit{SI Appendix}, Fig. S3.
    Self and cross-correlation functions  are shown in Fig. \ref{Fig2}A, 
    the corresponding response functions $J(t)$, obtained according to  Eq. \ref{Eq6}, are shown in Fig. \ref{Fig2}B.
    All self and cross-response functions smoothly decay to zero. The relaxation time, defined as the largest decay time $\tau_\mathrm{max}$ of a multi-exponential fit of $J$ (\textit{SI Appendix}, section 3), 
 is found to be the fastest for twist ($\tau_\mathrm{max} = 2.6$ ns), followed by shift ($\tau_\mathrm{max} = 7.1$ ns) and splay ($\tau_\mathrm{max} = 9.4$ ns).
        As expected, for each mode the two self response functions are greater than the cross response at all times. 
 The real and imaginary parts of the Fourier-transformed response functions, Re $\Tilde{J}(\omega)$ and Im $\Tilde{J}(\omega)$,
 are shown in Fig. \ref{Fig2}C,D. There is a distinct low-frequency plateau/peak in the real/imaginary part of the twist response.  
    To obtain  analytical representations, we fit multi-Debye functions (dashed lines) to the Fourier-transformed
    response functions $\Tilde{J}(\omega)$ in Fig. \ref{Fig2}C,D (details are provided in \textit{SI Appendix}, sections 2 and  3).
    Inverse Fourier transforms of the fit functions (dashed lines) reproduce the time-domain response
    functions, $J(t)$, data shown in Fig. \ref{Fig2}B very well. \par
\begin{figure}[t]
    \centering
    \includegraphics[width=0.42\textwidth]{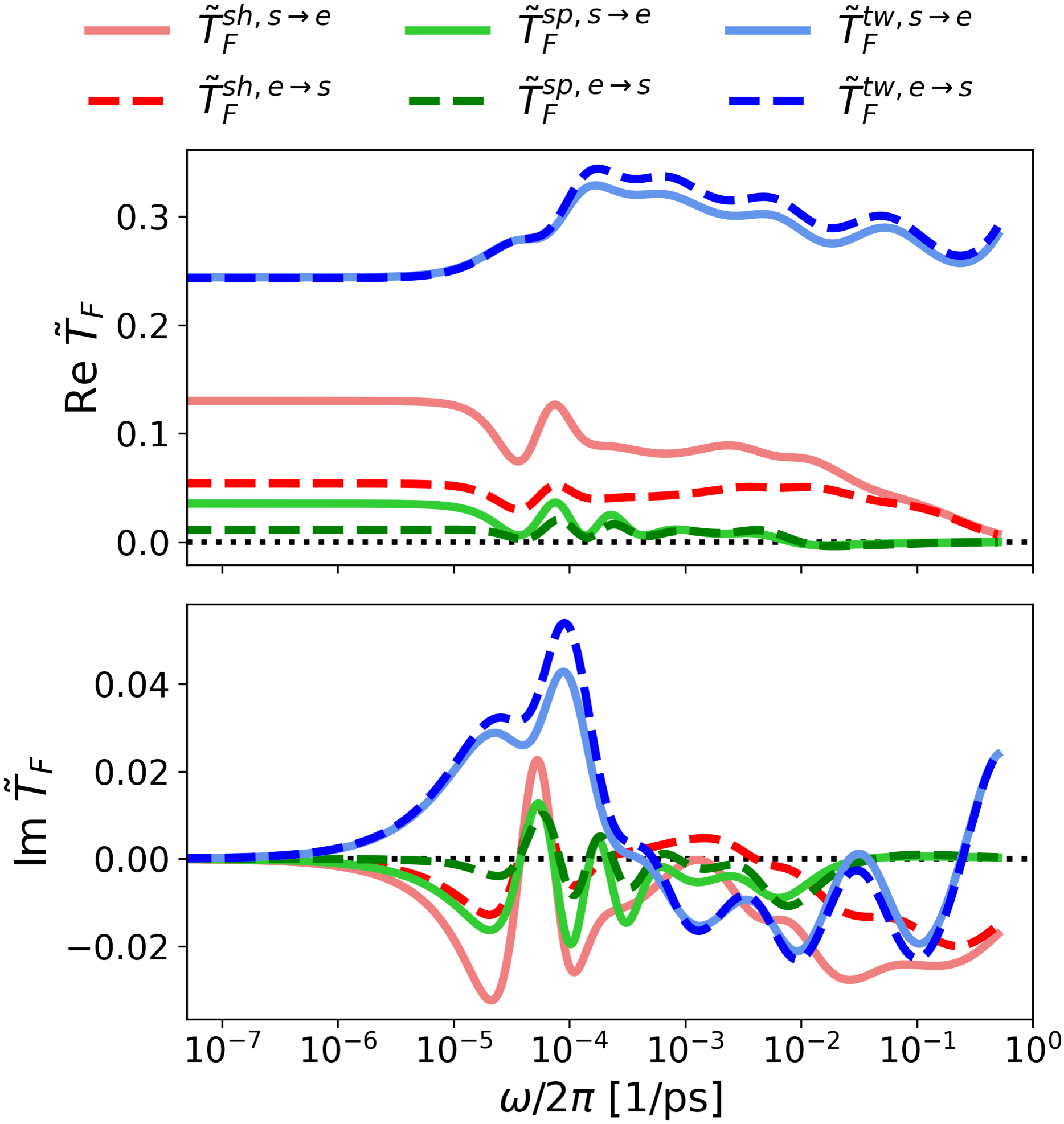}
    \caption{\label{Fig3} (top) Real and (bottom) imaginary part of force transmit functions $\Tilde{T}_\mathrm{F}$ for the shift (sh), splay (sp), and twist (tw) modes of the CC obtained using analytical representations for the self and cross-response functions (details in \textit{SI Appendix}, section 3
    ) according to  Eq. \ref{Eq3} . The sensor-to-effector (s $\rightarrow$ e) and effector-to-sensor (e $\rightarrow$ s)  transmit functions are shown as solid and dashed lines, respectively.} 
    \end{figure}

{}
    Force transmit functions, $\Tilde{T}_\mathrm{F}(\omega)$, obtained from the fitted $\Tilde{J}_\mathrm{self}(\omega)$ and $\Tilde{J}_\mathrm{cross}(\omega)$ using Eq. \ref{Eq3} are shown
    for all three different signaling modes in Fig. \ref{Fig3}. $\Tilde{T}_\mathrm{F}(\omega)$ quantifies
    the system's response to all possible excitation frequencies. 
         We find that for the entire frequency range, force transmission through the twist mode (blue) is the highest, followed by the shift mode (red) and the splay mode (green).
    From Re $\Tilde{T}_\mathrm{F}(\omega)$ in Fig. \ref{Fig3}, it is also evident that no force transmission is possible via the shift and splay mode for an input signal of frequency $> 400$ GHz ($\times 10^{-3} \mathrm{ps}^{-1}$) and $> 10$ GHz, respectively. These  cutoff frequencies are similar to  the water Debye mode at a  frequency of about  $\simeq 20$ GHz \cite{rinne2015impact},
    which suggests that the damping is partially due to the coupling to the hydration water. 
    Except for the   twist mode, the transmit functions 
  are generally asymmetric, i.e., the sensor-to-effector side transmit function $\Tilde{T}_\mathrm{F}^\mathrm{s\rightarrow e}(\omega)$ and the effector-to-sensor side transmit function $\Tilde{T}_\mathrm{F}^\mathrm{e\rightarrow s}(\omega)$ are different. 
   It should be noted that  $\Tilde{T}_\mathrm{F}(\omega)$ presented here characterizes the transmit  properties of an isolated parallel CC. 
   The effects of added sensor and effector protein modules can straightforwardly be obtained from the sensor and effector response functions using our 
   previously developed convolution 
   theory \cite{von2013convolution}. \par{}
\subsection{Signal transmission in the time domain}
\begin{figure*}[t]
    \centering
    \includegraphics[width=1.00\textwidth]{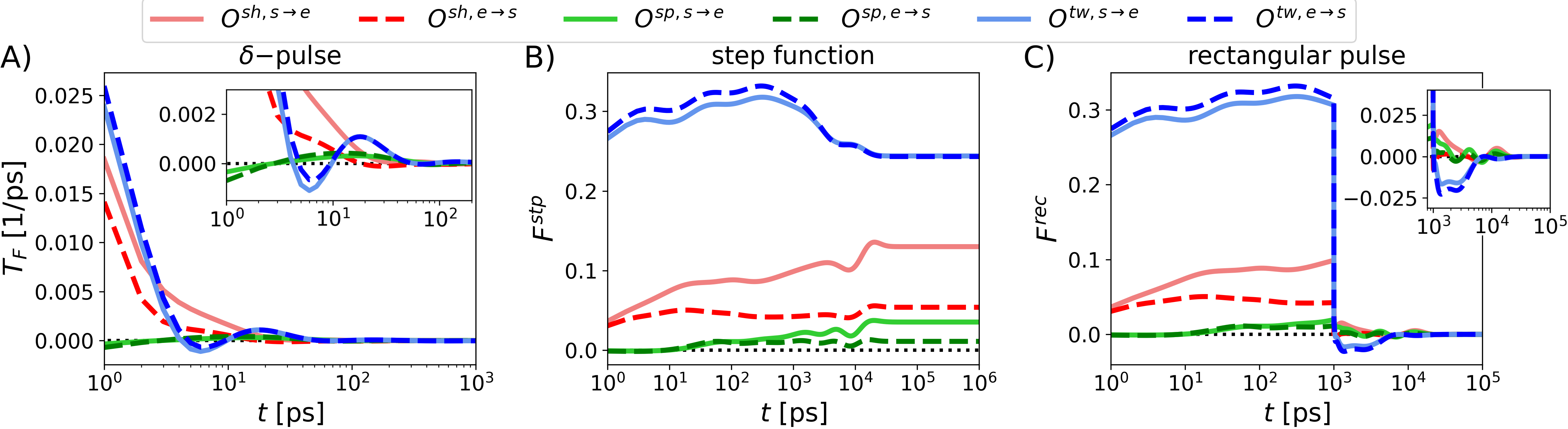}
    \caption{\label{Fig4} Transmitted force profiles for the different modes of the CC for different input force signals: 
    A)  $\delta$ pulse, B) step function, and C) rectangular pulse of width $\tau = 1$ ns. Insets in panel A and C represent zoomed-in force profiles.
    The force profiles in B and C are rescaled by the input force strength.
    } 
    \end{figure*}

For practical purposes,
 signal transmission in the  time domain needs to be characterized. 
 The force transmit  function in the time domain, $T^\mathrm{s \rightarrow e}_\mathrm{F}(t)$, 
 is given by the inverse Fourier transform of $\Tilde{T}^\mathrm{s \rightarrow e}_\mathrm{F}(\omega)$
and  describes  the transmission of a $\delta$-function input force signal $F_\mathrm{s}(t)$ at the sensor side.
      Once $T_\mathrm{F}^\mathrm{s \rightarrow e}(t)$ is known, 
      the transmitted force  $F_\mathrm{e}(t)$ at the effector side due to an arbitrary  input force  signal  $F_\mathrm{s}(t)$ can be obtained via  convolution 
    \begin{eqnarray}\label{Eq7}
        F_\mathrm{e}(t) = \int_{-\infty}^{t} T_\mathrm{F}^\mathrm{s \rightarrow e}(\tau) F_\mathrm{s}(t-\tau) d \tau.
    \end{eqnarray} 
    $T_\mathrm{F}(t)$ for the different signaling modes are shown in Fig. \ref{Fig4}A, 
    the method by which we obtain analytical representations for $T_\mathrm{F}(t)$ from $\Tilde{T}_\mathrm{F}(\omega)$
    is explained  in \textit{SI Appendix}, section 4. It is seen that the transmitted force signals for all the different modes decay rather quickly within 100 ps. \par{}
In reality, signals are not transmitted via infinitely short $\delta$-pulses but rather by pulses of finite duration. 
How efficiently  the CC linker transmits such an input force signal depends on how quickly it responds to a suddenly imposed force
and how quickly it relaxes back  to its equilibrium state after the removal of force. 
To understand theses complex dynamics, we discuss  the transmitted output forces for input force step  and  rectangular force pulse, $F_\mathrm{e}^\mathrm{stp}(t)$ and $F_\mathrm{e}^\mathrm{rec}(t)$, 
both obtained  from the convolution integral Eq. \ref{Eq7}, for further details see \textit{SI Appendix}, section 4. 
  For a  force switched on at $t = 0$, $F_\mathrm{e}^\mathrm{stp}(t)$ is shown in Fig. \ref{Fig4}B for the different modes. 
 The transmitted force plateau values are reached for the different modes after 30--50 ns, which is around 2 orders of magnitude faster than the 
 reported experimental timescale of 2 $\mu$s associated with light-induced conformational transitions within the sensor module \cite{berntsson2017sequential}. 
Signal transmission by all three modes would hence be sufficiently fast to not become time-limiting.
 The plateau value of $F_\mathrm{e}^\mathrm{stp}$ is the highest for the twist mode, followed by shift and splay modes.
In Fig. \ref{Fig4}C we present the transmitted force  $F_\mathrm{e}^\mathrm{rec}(t)$  for  a rectangular force pulse  signal of duration $\tau = 1$ ns that is switched on at  $t = 0$.
It is seen that the transmitted force through each mode decay to zero within 40 ps after the removal of applied force.
Results for different durations of the rectangular force pulse of  $\tau = 10^{-2}, 10^{-1}, 10^1, 10^2$ ns are presented  in \textit{SI Appendix} Fig. S4. \par{}
    \subsection{Robustness with respect to mutations}
\begin{figure*}[]
    \centering
    \includegraphics[width=0.945\textwidth]{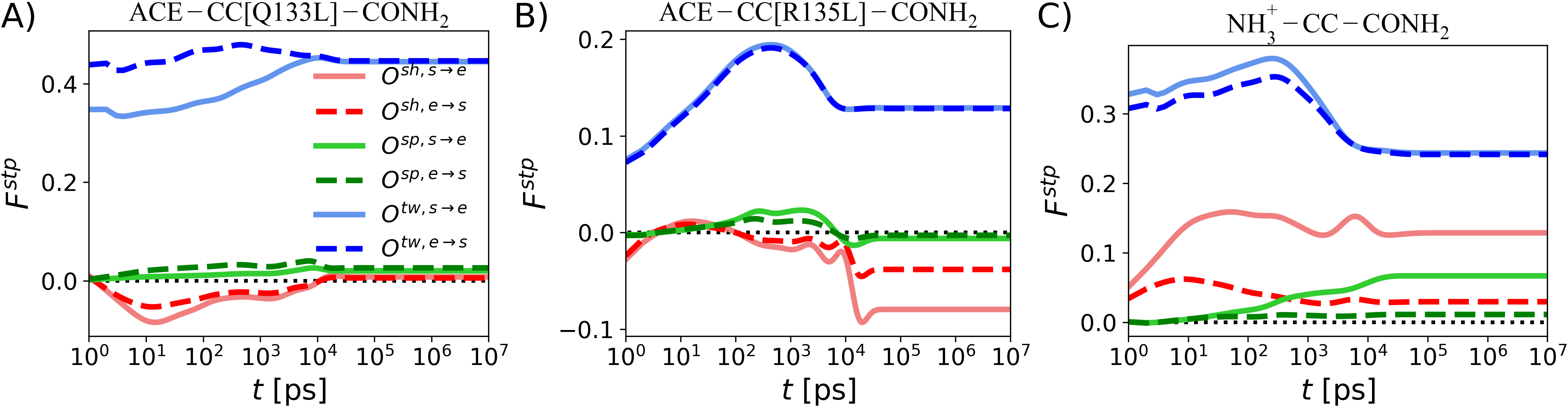}
    \caption{\label{Fig5} Comparison of transmitted force profiles for an input step force for modified CC systems: 
    A,B) CC with different single-point mutations (Q133L or R135L) and C) CC with charged N-termini. All force profiles are rescaled by the input force strength.
    For a comparison with the wild-type CC with charge-neutral N and C-termini, see Fig. \ref{Fig4}B.}
    \end{figure*}

    It has been experimentally demonstrated that single-point mutations within the CC reduce the signal response of the blue-light-regulated  histidine kinase YF1 \cite{diensthuber2013full, gleichmann2013charting}.  To study this in our framework, we perform MD simulations of two different experimentally studied  mutants,
     Q133L, where Gln$^{133}$ is replaced by Leu, and R135L, where  Arg$^{135}$ is replaced by Leu.
We find that these single-point mutations do not affect the overall coiled-coil conformation, as seen from their different structural order parameter values compared with that of the wild-type CC in \textit{SI Appendix}, Fig. S1. 
    However, the dynamics and signaling response of these two mutants are completely different from each other and from the wild-type CC.
 For the mutant Q133L  in Fig. \ref{Fig5}A, the step force transmission for the shift and splay modes are negligible.
 Interestingly, the  plateau value of the twist-mode in Fig. \ref{Fig5}A  is larger than that of the wild-type CC  in Fig. \ref{Fig4}B.
 Based on the absence of signaling for Q133L in experiments \cite{gleichmann2013charting}, this suggests that the signaling in the histidine kinase is not connected to the twist mode.
 For the mutant R135L  in Fig. \ref{Fig5}B, 
 the plateau value of the shift-mode transmission is finite but negative, the splay-mode step transmission is much reduced, 
 and the plateau value of the twist-mode transmission is almost half in comparison to the wild-type CC, in experiments the signaling activity is reduced but not absent \cite{gleichmann2013charting}.
In addition, structural characterization by electron paramagnetic resonance spectroscopy and X-ray solution scattering has revealed that light induces a splaying apart of the sensor domains and hence of the N-termini of the CC \cite{berntsson2017sequential, berntsson2017time}.
 Thus, by comparison with the experimental findings, we conclude that the 
 signaling mode in the histidine kinase is predominantly of the splay type. 
 Berntsson \textit{et al.} \cite{berntsson2017sequential} have experimentally
  observed superhelical coiling of the CC subsequent to its light-induced splaying, which suggests
  coupling between splay and twist deformations \cite{moglich2019signal}. We quantify the coupling
  between different deformation modes by calculating the Pearson correlation coefficients 
  (\textit{Materials and Methods}) as summarized in Table \ref{Tab1} and indeed find significant splay-twist coupling which could explain the experimentally seen twisting by coupling to splay.
 We note that our simulations reveal that the twist mode is most stable with respect to  mutations,
 so it is conceivable that the CC linker might  in a different biological context also function as a twist transmitter. \par{} 
     \begin{table}[h]
      \begin{center}
        \caption{Correlation coefficients between different deformation modes for the same ends (sensor and effector) and for different ends (cross).}
         \label{Tab1}
         \begin{tabular}{|l|c|c|c|}
           \hline
           type & sh-sp & sh-tw & sp-tw\\
           \hline
           sensor & -0.021 & -0.012 & -0.012\\
           effector & 0.26 & -0.011 & 0.004\\
           cross & -0.057 & -0.006 & 0.006\\
           \hline
         \end{tabular}
       \end{center}
     \end{table}
       \subsection{Transmission asymmetry}
Asymmetry could possibly be important  for the efficient information transfer from the sensor to the effector side. 
To look into this, we introduce the \textit{rectification factor} $\gamma$ as the ratio of sensor-to-effector ($\mathrm{s \rightarrow e}$) and
effector-to-sensor ($\mathrm{e \rightarrow s}$) step-force transmission profile plateau values: 
    \begin{eqnarray*}
        \gamma = \lim_{t\to\infty} \frac{F_\mathrm{e}^\mathrm{stp}(t)}{F_\mathrm{s}^\mathrm{stp}(t)}.    
    \end{eqnarray*}
  From the results in Fig. \ref{Fig4}B we conclude that rectification for the splay mode is the highest, $\gamma^\mathrm{sp} = 3.2$, 
 followed by the shift mode, $\gamma^\mathrm{sh} = 2.4$. In contrast, no rectification is observed for the twist mode, i.e.,  $\gamma^\mathrm{tw} = 1$. 
 These results can be rationalized by the fact that the rectification factor is given by the ratio of the real parts of the zero-frequency self responses 
 for the sensor and the effector side, see \textit{SI Appendix}, Fig. S3C. 
  To study the relation between the rectification factor and the structural asymmetry in more detail, 
we  introduce an additional asymmetry between the sensor and  effector ends of the CC by uncapping the sensor-side $\alpha$-helix termini,
which thereby become positively charged at neutral pH, resulting in the structure  NH$_3^+$-CC-CONH$_2$
 (note that the  results presented in Figs. \ref{Fig1}--\ref{Fig4} are obtained  for the CC linker with  charge-neutral end groups:  ACE-CC-CONH$_2$).
    Though the step-force transmission profiles for ACE-CC-CONH$_2$ and NH$_3^+$-CC-CONH$_2$ are qualitatively the same,
    as follows by comparing  Figs. \ref{Fig4}B and \ref{Fig5}C, we observe a pronounced difference of the  plateau values for  the 
$\mathrm{s \rightarrow e}$ and $\mathrm{e \rightarrow s}$
    transmission  for the shift and splay modes of NH$_3^+$-CC-CONH$_2$. However, its
    $\mathrm{s \rightarrow e}$ and $\mathrm{e \rightarrow s}$ plateau values for the  twist transmission 
     are the same and remain unaffected in comparison to the charge neutral-termini system ACE-CC-CONH$_2$.
     We thus find that the rectification factor can be tuned by changing the chemical structure of the  sensor and effector terminal groups. 
     It is tempting to relate the $\gamma$ values for the different modes to their functional relevance, e.g., the highest value for $\gamma^\mathrm{sp}$ found here and the light-induced splaying of the CC linker observed in experiments \cite{berntsson2017sequential, berntsson2017time}.

       \section*{Discussion and Conclusion}

  We introduce the theoretical framework to quantify the signal transmission between two distinct sites of a protein in terms of  the associated
  self  and  cross response functions. The response functions are via  the fluctuation--dissipation theorem related to   equilibrium time-correlation functions, 
  which  can be generated from MD simulations but also from  single-molecule experiments. 
  Note that in experiments, trajectories of separation coordinates typically include effects due to the  coupling to measurement devices, 
  which can  be filtered out by  using dynamic deconvolution theory \cite{hinczewski2010deconvolution, von2012auto}.
  The displacement transmit function 
  relates the correlations between two sites and the fluctuations at the site
  at which the input signal is applied, it thus quantifies the ratio of the output to the input signal and 
  thus conveys more useful information than the often considered dynamic cross-correlation. \par{}
  We apply  our  theoretical framework to the CC linker from the bacterial signaling protein YF1 and
  show that all three deformation modes, twist, shift and splay, achieve signal transfer 
  from the sensor to the effector end of the CC. 
  Although twist is in principle a better signal transmission mode, it does not conserve 
  angular momentum and will therefore lead to a rotation of the sensor domain, this is probably
  why nature is not using it, at least for this protein construct. The experimentally observed splaying
  followed by superhelical coiling of the CC upon triggering \cite{berntsson2017sequential} are expected
  due to a coupling between splay and twist deformation modes.
  Our analysis of simulation data for the wild-type CC and two important single-point mutants
  \cite{diensthuber2013full} suggests that splay is actually the signaling mode realized in the
  experimentally studied histidine kinase \cite{berntsson2017sequential, berntsson2017time}. \par{}
   Previous experiments have indicated that the length of the CC linker, not only the actual linker
   sequence, is instrumental in determining the response to light signals \cite{ohlendorf2016library}.
   Our framework is directly applicable to study the CC length-dependent signaling.     
   Moreover, our method will be useful to understand the dynamics of activation pathways of 
    other cell signaling proteins with complex topology, e.g., G-protein-coupled receptors 
    \cite{thal2018structural, hilger2018structure, rodriguez2020gpcrmd}, which are the most frequent targets of drugs due to their involvement in diverse physiological processes.   
    In this context it should be kept in mind that signal transmission through general protein networks can be predicted from  the response functions of individual components
   by repeated application of convolution relations for serial and parallel connections \cite{hinczewski2010deconvolution, von2013convolution}. 
   Our study, thus, provides a way forward to relate atom-level protein dynamics to large-scale intermolecular communications of biological  signaling networks. \par{}
 Our theory is formulated at the linear-response level and thus is scale-invariant with respect to the input signal amplitude. 
  In order to obtain the signal threshold beyond which the signal strength surpasses the noise background, 
  one  needs to compare the signal strength with the root-mean-square  of the fluctuating force or displacement, 
  similar to the definition of the signal-to-noise ratio in information theory \cite{mackay2003information}. 
 We have in this paper only considered signaling between identical deformation 
  modes at the two ends, i.e. twist to twist, splay to splay and shift to shift, off-diagonal signaling
  modes might be relevant experimentally and will be considered in future work. \par{}

\section*{Materials and Methods}  
    \subsection*{Models and force-field parameters} 
    From the crystal structure of the whole signal transducing protein unit (PDB ID: 4GCZ), sensor (N-terminal) and effector (C-terminal) modules are deleted to obtain the structure of the CC linker \cite{diensthuber2013full}. The CC is composed of two parallel $\alpha$-helices each containing the same 23 residues ([126]Ile-Thr-Glu-His-Gln-Gln-Thr-Gln-Ala-Arg-Leu-Gln-Glu-Leu-Gln-Ser-Glu-Leu-Val-His-Val-Ser-Arg[148]). The CC is simulated in a rhombic dodecahedron box of volume 227 nm$^3$ filled with 7135 water molecules (and counterions needed to neutralize the system). CHARMM36m protein force field parameters \cite{huang2017charmm36m}, the  TIP3P water model \cite{jorgensen1983comparison, mackerell1998all} and ion parameters from Ref. \cite{venable2013simulations} are used. Four different systems with changes of N- and C-termini capping groups and/or a mutated residue are considered: NH$_3^+$-CC-CONH$_2$, ACE-CC-CONH$_2$, ACE-CC[Q133L]-CONH$_2$, ACE-CC[R135L]-CONH$_2$. The acetyl (ACE) cap, CH$_3$-CO-, is used at the N-terminal and the ``-CONH$_2$" group is used at the C-terminal Arg, to simulate charge neutral termini.  The two mutated systems are selected from the study by Gleichmann \textit{et al.}  \cite{gleichmann2013charting}. \par{}
    \subsection*{MD simulation details} 
    For each system, a simulation is performed for 20 $\mu$s in the $NpT$ ensemble at temperature $T=300$ K and pressure $p=1$ bar with periodic boundary conditions using Gromacs 2020.1 \cite{abraham2015gromacs}. The stochastic velocity rescaling thermostat \cite{bussi2007canonical} with a time constant $\tau_{T}=0.1$ ps is used to control temperature, while for pressure control an isotropic Parrinello--Rahman barostat \cite{parrinello1981polymorphic} is used with a time constant $\tau_{p}=2$ ps and compressibility $\kappa=4.5\times10^{-5}$ bar$^{-1}$. The LINCS algorithm \cite{hess1997lincs} is used to constrain the bonds involving hydrogen atoms, allowing a timestep $\Delta t=2$ fs. Electrostatic interactions are computed using the particle mesh Ewald method \cite{darden1993particle} with a real-space cutoff distance of 1.2 nm, while van der Waals interactions are modeled using Lennard-Jones potentials with a cutoff distance of 1.2 nm where the resulting forces smoothly switch to zero between 1 nm to 1.2 nm. For the data analysis, simulation trajectories are saved every 1 ps. Images are rendered using the visual molecular dynamics (VMD) software \cite{humphrey1996vmd}. Analysis is performed using in-house developed codes and Gromacs analysis modules \cite{abraham2015gromacs}. Correlation functions  for observable $A(t)$ and $B(t)$ are calculated as 
    \begin{eqnarray*}
       C(\tau) = \frac{1}{L-\tau} \int_{0}^{L-\tau} A(t)B(t+\tau)dt = \langle A(t)B(t+\tau) \rangle,
    \end{eqnarray*}
    where $L$ is the trajectory length.
    \subsection*{Protein rotational relaxation time} 
    The rotational relaxation time $\tau_\mathrm{r}$ of an object is related to the rotational diffusion coefficient as $D_\mathrm{r} = 1/2\tau_\mathrm{r}$ and is estimated from the Stokes' rotational diffusion coefficient $D_\mathrm{r} = k_\mathrm{B}T / 8 \pi \eta R_\mathrm{h}^3$ \cite{landau2013fluid}. Using the viscosity of the medium as that of water, $\eta= 8.9 \times 10^{-1}$ Pa$\cdot$s, and the hydrodynamic radius as half of the largest length scale of the full-length protein YF1, $R_\mathrm{h} = 7$ nm, we obtain $\tau_\mathrm{r} = 2\ \mu$s which exceeds the transmission relaxation times by far.
  \subsection*{Coupling between deformation modes} 
  Pearson correlation coefficients between two different deformation modes are obtained as 
     \begin{eqnarray*}
     	R_{ij} = \frac{\langle (x_{i} - \langle x_{i} \rangle) (x_{j} - \langle x_{j} \rangle) \rangle}{\sqrt{\langle (x_{i} - \langle x_{i} \rangle)^2\rangle} \sqrt{\langle (x_{j} - \langle x_{j} \rangle)^2\rangle}}, 
     \end{eqnarray*}
     where $\langle \cdot \rangle$ denotes the time average and $i, j$ refers to shift, splay or twist at the same ends or at different ends of the CC. The correlation coefficients $R_{ij}$ for $i \neq j$ (as $R_{ii} = 1$) are summarized in Table \ref{Tab1}.
    \subsection*{Data, Materials, and Software Availability} 
    All study data are included in the main text and/or \textit{SI Appendix}. Derivations and additional figures that support the findings of this study are included in \textit{SI Appendix}. 

\section*{Acknowledgements}
We acknowledge funding by the European Research Council (ERC) Advanced Grant NoMaMemo Grant No. 835117, the Infosys foundation and computing time on the HPC cluster at ZEDAT, FU Berlin.

%

\end{document}


\title{{\huge Supporting Information for} \\ Quantifying the efficiency of principal signal transmission modes  in proteins}

\author{Anil Kumar Sahoo, Richard Schwarzl, Markus S. Miettinen, Roland R. Netz}


\maketitle

\clearpage

\section*{Supporting Infromation Text}
\subsection{\label{App1}Order parameters for characterizing the coiled-coil (CC) structure}
    \textit{Fraction of native contacts}, $Q$, is defined as \cite{best2013native}:
    \begin{eqnarray*}
        Q(t) = \frac{1}{N_c}\sum_{(i,j)}\frac{1}{1+\exp{\lambda[r_{ij}(t)-\nu r_{ij}^0]}},
    \end{eqnarray*}
    where the sum runs over $N_c$ pairs of native contacts $(i, j)$, $r_{ij}(t)$ is the distance between $i$ and $j$ at time $t$, $r_{ij}^0$ is the distance between $i$ and $j$ in the native structure of the protein, $\lambda$ ($= 5$ \AA$^{-1}$) is a smoothening parameter, and $\nu$ ($= 1.8$) takes care of the fluctuations during contact formation. Here, the list of native contact pairs $(i, j)$ is constructed by considering all pairs of heavy atoms belonging to residues $R_i$ and $R_j$ such that $|R_i - R_j| > 3$ and the distance between $i$ and $j$ is less than 4.5 \AA.\\ \\
    \textit{Root-mean-square deviation}, RMSD, in distances is defined as
    \begin{eqnarray*}
        RMSD(t) = \sqrt{\frac{1}{N}\sum_{i=1}^{N}|r_i(t)-r_i^0|^2},
    \end{eqnarray*}
    where $N$ is the total number of heavy atoms of the CC, $r_i(t)$ is the position of the $i^\mathrm{th}$ atom at time $t$, and $r_i^0$ is the position of the same atom in the native structure of the CC. \\ \\
    \textit{Secondary structure}, SS, contents of the CC are calculated by the STRIDE algorithm \cite{frishman1995knowledge} implemented in the VMD software  \cite{humphrey1996vmd}. The fraction of SS content is defined as the ratio of the number of residues of the protein involved in $\alpha$-helical SS formation at time $t$ to that in the native structure of the CC. \par{}
 
\subsection{\label{App2}Force transmission through a viscous bead--spring model}
    To illustrate force transmission through a model signal transmitter, we consider a simple viscoelastic model: two beads with different mobilities $\mu_\mathrm{s}$ (sensor) and $\mu_\mathrm{e}$ (effector) that are connected by a harmonic spring  of stiffness $k$, similar to  Fig. 1B in the main text \cite{hinczewski2010deconvolution}. 
    Under the application of oscillating forces with amplitudes $f_\mathrm{s}(\omega)$ and $f_\mathrm{e}(\omega)$ to the sensor and the effector, respectively,
     the Fourier-transformed sensor velocity is given by $-i\omega \Tilde{x}_\mathrm{s}(\omega) = \mu_\mathrm{s}(f_\mathrm{s}(\omega) + 
     k[\Tilde{x}_\mathrm{e}(\omega)-\Tilde{x}_\mathrm{s}(\omega)])$, 
     whereas the effector velocity is given by $-i\omega \Tilde{x}_\mathrm{e}(\omega) = \mu_\mathrm{e}(f_\mathrm{e}(\omega) +
      k[\Tilde{x}_\mathrm{s}(\omega)-\Tilde{x}_\mathrm{e}(\omega)])$. Using the definitions of linear response functions from Eq. 1 in the main text, we obtain
    \begin{eqnarray}
    \begin{split}
        \Tilde{J}_\mathrm{self}^\mathrm{s} = \frac{\mu_\mathrm{s}(\omega + i\mu_\mathrm{e} k)}{\omega(\mu k -i\omega)}, \ \ \
        \Tilde{J}_\mathrm{self}^\mathrm{e} = \frac{\mu_\mathrm{e}(\omega + i\mu_\mathrm{s} k)}{\omega(\mu k -i\omega)}, \ \ \
        \Tilde{J}_\mathrm{cross} = \frac{i\mu_\mathrm{s}\mu_\mathrm{e} k}{\omega(\mu k -i\omega)}, 
    \end{split}
    \end{eqnarray}
    where $\mu = \mu_\mathrm{s} + \mu_\mathrm{e}$. The force transmission from the sensor to the effector side is determined according to  Eq. 2 in the main text as
    \begin{eqnarray}
        \Tilde{T}_\mathrm{F}^\mathrm{s \rightarrow e} (\omega) = 
        \frac{\Tilde{J}_\mathrm{cross} (\omega)}{\Tilde{J}_\mathrm{self}^\mathrm{e} (\omega)}
        = \frac{\mu_\mathrm{s} k}{(\mu_\mathrm{s} k -i\omega )}.
    \end{eqnarray}
    Note that $\Tilde{T}_\mathrm{F}^\mathrm{s \rightarrow e}$ is independent of the effector side bead mobility, and
     at zero frequency, which corresponds to the long-time limit, the force transmission is given by
      $\Tilde{T}_\mathrm{F}^\mathrm{s \rightarrow e} = 1$. The derived force transmit function has the same form as the Debye  function. 
    For more realistic force transducers, such as proteins, $\Tilde{T}_\mathrm{F}$ can be expressed as a sum of Debye functions that reflect the normal modes of a protein.

\subsection{\label{App3}Fitting procedure for the analytical representation of a response function}
    We consider multi-exponential functions
    for fitting the response function $J(t)$ according to 
    \begin{equation}{\label{Eq1_A3}}
        J(t) = \theta(t) \sum_{k=1}^{N} c_k e^{-t/\tau_k} = \sum_{k=1}^{N} g_k(t),
    \end{equation}
    where $\theta$ is the Heaviside step function given by
    \begin{equation*}
        \theta(t)=    
        \begin{cases}
        1 & t\geq0\\
        0 & t<0.
       \end{cases} 
    \end{equation*}
    The Fourier transform of a single-sided  exponential is the  Debye function given by
    \begin{equation}{\label{Eq2_A3}}
        \Tilde{g}_k(\omega)= \frac{c_k} {\tau_k^{-1}+i\omega} = \frac{c_k\tau_k}{1+\tau_k^2\omega^2} -i\frac{c_k\tau_k^2\omega}{1+\tau_k^2\omega^2}.
    \end{equation}    
    The real part of the fit function is  given by
    \begin{equation}{\label{Eq3_A3}}
        \Re\ [\Tilde{J}(\omega)] = \Re \big[\sum_{k=1}^{N} \Tilde{g}_k(\omega)\big]= \sum_{k=1}^{N} \frac{c_k\tau_k}{1+\tau_k^2 \omega^2}
    \end{equation}
    and the imaginary part is given by
    \begin{equation}{\label{Eq4_A3}}
        \Im\ [\Tilde{J}(\omega)] = \Im \big[\sum_{k=1}^{N} \Tilde{g}_k(\omega)\big]= -\sum_{k=1}^{N} \frac{c_{k}\tau_{k}^{2}\omega}{1+\tau_k^2\omega^2}.
    \end{equation}
    Eqs. \ref{Eq3_A3},\ref{Eq4_A3} are used to simultaneously fit the real and imaginary part of the Fourier-transformed response functions. Logarithmically-spaced frequency data are taken for the fitting. The time constants $\tau_k$ in Eq. \ref{Eq1_A3}  are restricted to positive values only. The number of Debye  functions, $N$, used  for fitting the self and cross-response functions for the different systems are given in Table \ref{Tab1_A3}. Such fits for the twist mode and the shift and splay modes of the ACE-CC-CONH$_2$ system are shown in Figs. 2C,D in the main text and \ref{Fig-S3}C,D, respectively.  
    Importantly, we find that the parameters $c_k$ and $\tau_k$  obtained from the frequency-domain fitting accurately reproduce the simulated time-domain response functions, validating the robustness of our fitting procedure (see Figs. 2B in the main text and \ref{Fig-S3}B).
    \begin{table}[h!]
      \begin{center}
        \caption{Number of Debye relaxation functions $N$ used for fitting self and cross-response functions for the different systems.}
        \label{Tab1_A3}
        \begin{tabular}{|l|c|c|}
          \hline
          System & $N (\Tilde{J}_\mathrm{self})$ & $N (\Tilde{J}_\mathrm{cross})$ \\
          \hline
          ACE-CC-CONH$_2$ & 10 & 10\\
          NH$_3^+$-CC-CONH$_2$ & 7 & 10\\
          ACE-CC[Q133L]-CONH$_2$ & 7 & 10\\
          ACE-CC[R135L]-CONH$_2$ & 7 & 10\\
          \hline
        \end{tabular}
      \end{center}
    \end{table}
    %
\subsection{\label{App4}Time-domain transmit function from the analytical form of $\Tilde{T}_\mathrm{F}(\omega)$ via partial fraction decomposition}
    The time-domain force transmit function,  $T_\mathrm{F}(t)$, 
     corresponds to the  system response to a $\delta$-function input pulse  and  is given by the inverse Fourier transform of the Fourier-domain transmit function, $\Tilde{T}_\mathrm{F}(\omega)$, according to
 $T_\mathrm{F}(t) = \frac{1}{2\pi} \int_{-\infty}^{+\infty} \Tilde{T}_\mathrm{F}(\omega) e^{i\omega t} d\omega$.
 Instead of using a numerical discrete Fourier transform, we  obtain $T_\mathrm{F}(t)$ from $\Tilde{T}_\mathrm{F}(\omega)$ analytically as follows. 
 Since all response functions $\Tilde{J}(\omega)$ are expressed as sums of Debye  functions (see Section \ref{App3}), the Fourier-transformed force transmit function is given by
    \begin{eqnarray}{\label{Eq1_A4}}
        \Tilde{T}_\mathrm{F}(\omega) = \frac{\Tilde{J}_\mathrm{cross} (\omega)}{\Tilde{J}_\mathrm{self} (\omega)} = \sum_{j=1}^{M} \frac{a_j}{\alpha_j + i\omega} \bigg/ \sum_{k=1}^{N} \frac{b_k}{\beta_k + i\omega},
    \end{eqnarray} 
    where $a_j$, $\alpha_j$, $b_k$, and $\beta_k$ are the fitting parameters. The above expression can be rewritten as a sum of Debye  functions using partial fraction decomposition as
    \begin{eqnarray}{\label{Eq2_A4}}
        \Tilde{T}_\mathrm{F}(\omega) = c_0 + \sum_{l=1}^{Q} \frac{c_l}{\gamma_l + i\omega},
    \end{eqnarray}
    where $c_0$, $c_l$, and $\gamma_l$ are uniquely determined  parameters. 
    Based on  Eq. \ref{Eq2_A4} the  analytical inverse Fourier transform is obtained as 
    \begin{eqnarray}{\label{Eq3_A4}}
        T_\mathrm{F}(t) = c_0\delta(t) + \theta(t)\sum_{l=1}^{Q} c_l e^{-\gamma_l t},
    \end{eqnarray}
    where $\theta(t)$ is the Heaviside step function.

    Once  the transmit function $T_\mathrm{F}(t)$ is known, the transmitted force  $F_e(t)$  at the effector site follows from 
    the input force  $F_s(t)$ at the sensor site by  convolution. 
    For a  step input force  signal $F_s(t) = F_0 \theta(t)$ we obtain 
    \begin{eqnarray*}
        F_\mathrm{e}^\mathrm{stp}(t) = F_0  \int_{-\infty}^{\infty} T_\mathrm{F}(t^{\prime}) \theta(t-t^{\prime}) dt^{\prime} 
        = F_0 \int_{-\infty}^{t} T_\mathrm{F}(t^{\prime}) dt^{\prime} \ \ \ \ \
        \end{eqnarray*}
    \begin{eqnarray}
        \implies F_\mathrm{e}^\mathrm{stp}(t)/F_0 =  c_0 \theta(t) +  \sum_{l=1}^{Q} \frac{c_l}{\gamma_l} (1 - e^{-\gamma_l t}).
    \end{eqnarray}
    Similarly, the rectangular pulse response is given by
    \begin{equation*}
        F_\mathrm{e}^\mathrm{rec}(t) = F_0  \int_{-\infty}^{\infty} T_\mathrm{F}(t^{\prime}) \Pi(t-t^{\prime}) dt^{\prime},
    \end{equation*} 
    where $\Pi(t)$ is a rectangular pulse of width $\tau$ defined as  
    \begin{equation*}
        \Pi(t)=    
        \begin{cases}
        0 & t < 0 \\
        1 & 0 \leq t \leq \tau \\
        0 & t > \tau .
        \end{cases} 
    \end{equation*}
    The above convolution integral is evaluated to be:
    \begin{eqnarray}
        F_\mathrm{e}^\mathrm{rec}(t) /F_0 = 
        \begin{cases}
        \int_{0}^{t} T_\mathrm{F}(t^{\prime})dt^{\prime} & 0 \leq t \leq \tau \\ \\
        \int_{t-\tau}^{t} T_\mathrm{F}(t^{\prime})dt^{\prime} & t > \tau
        \end{cases}  
        = 
        \begin{cases}
        c_0 + \sum_{l=1}^{Q} \frac{c_l}{\gamma_l} (1 - e^{-\gamma_l t}) & 0 \leq t \leq \tau \\ \\
        \sum_{l=1}^{Q} \frac{c_l}{\gamma_l} e^{-\gamma_l t} (e^{\gamma_l \tau} -1) & t > \tau .
        \end{cases}  
    \end{eqnarray}    
\clearpage

%
\begin{figure*}[]
\centering
\includegraphics[width=0.45\textwidth]{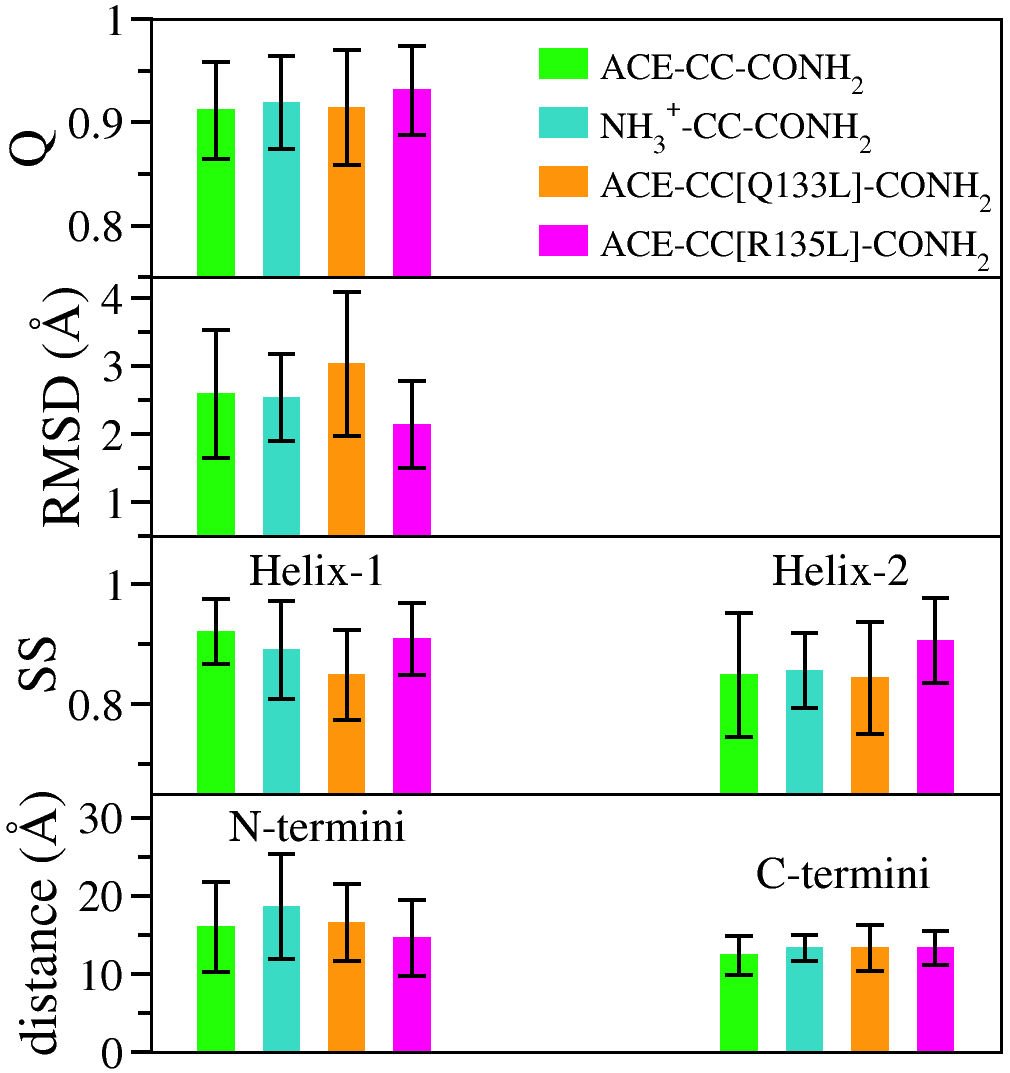}
\caption{\label{Fig-S1} Average values of different structural order parameters for the four simulated coiled-coil  systems: 
ACE-CC-CONH$_2$, NH$_3^+$-CC-CONH$_2$, ACE-CC[Q133L]-CONH$_2$, and ACE-CC[R135L]-CONH$_2$. See Section \ref{App1} for the definitions of the order parameters. The results demonstrate that all four systems are stable over the total simulation time of 20 $\mu$s each.} 
\end{figure*}
\clearpage
\begin{figure*}[]
\centering
\includegraphics[width=0.80\textwidth]{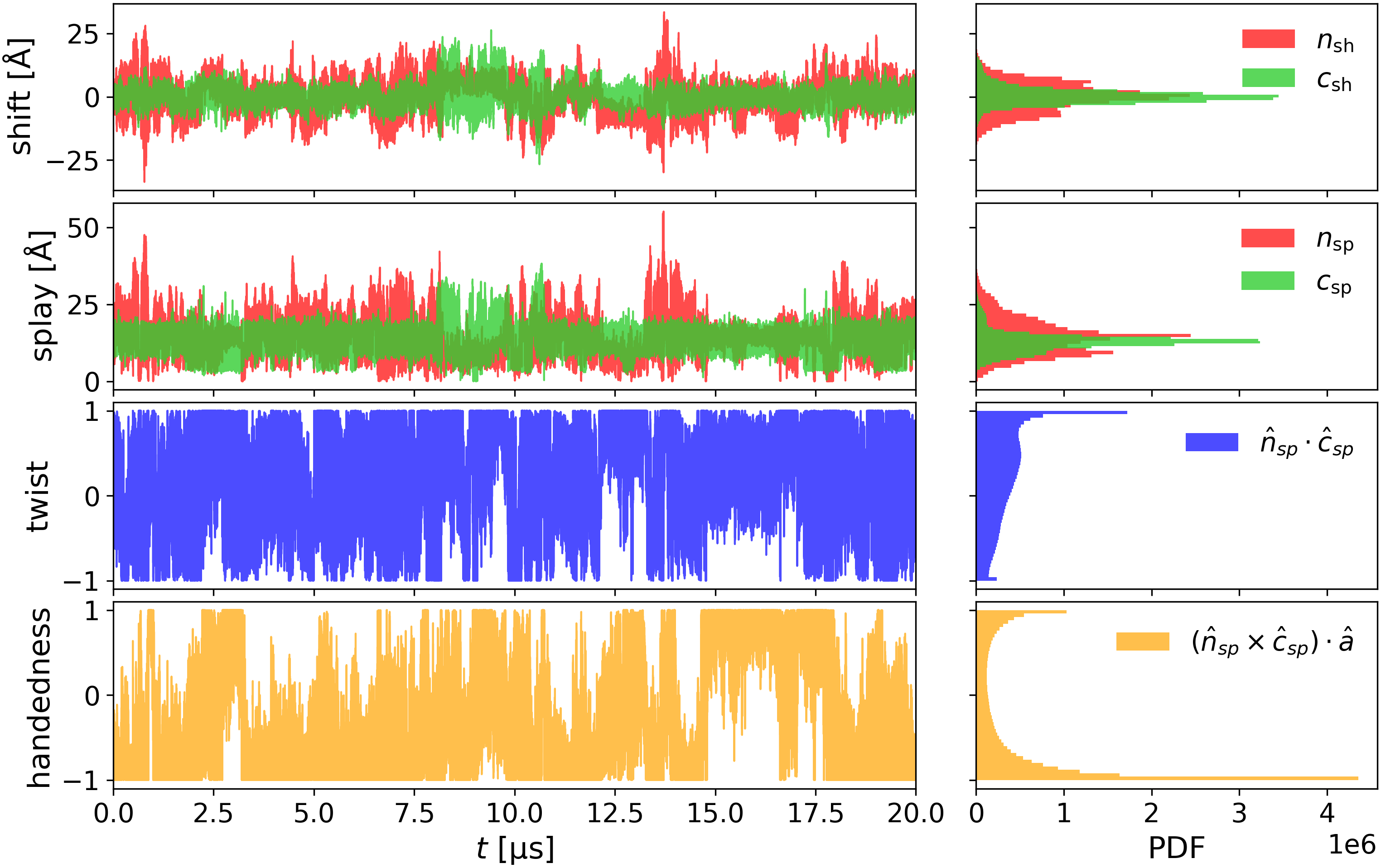}
\caption{\label{Fig-S2} Time series and corresponding probability distribution functions for the shift, 
the splay, the cosine of the twist angle between N and C termini, and the handedness of the rotation of the coiled coil along its long-axis $\vec{a}$, defined as $(\hat{n}_\mathrm{sp} \times \hat{c}_\mathrm{sp}) \cdot \hat{a}$. See Fig. 1 and the main text for the definitions of shift, splay, and the twist angle. Results are shown for the wild-type coiled coil with capped neutral termini (ACE-CC-CONH$_2$). } 
\end{figure*}
\clearpage
\begin{figure*}[]
\centering
\includegraphics[width=1.0\textwidth]{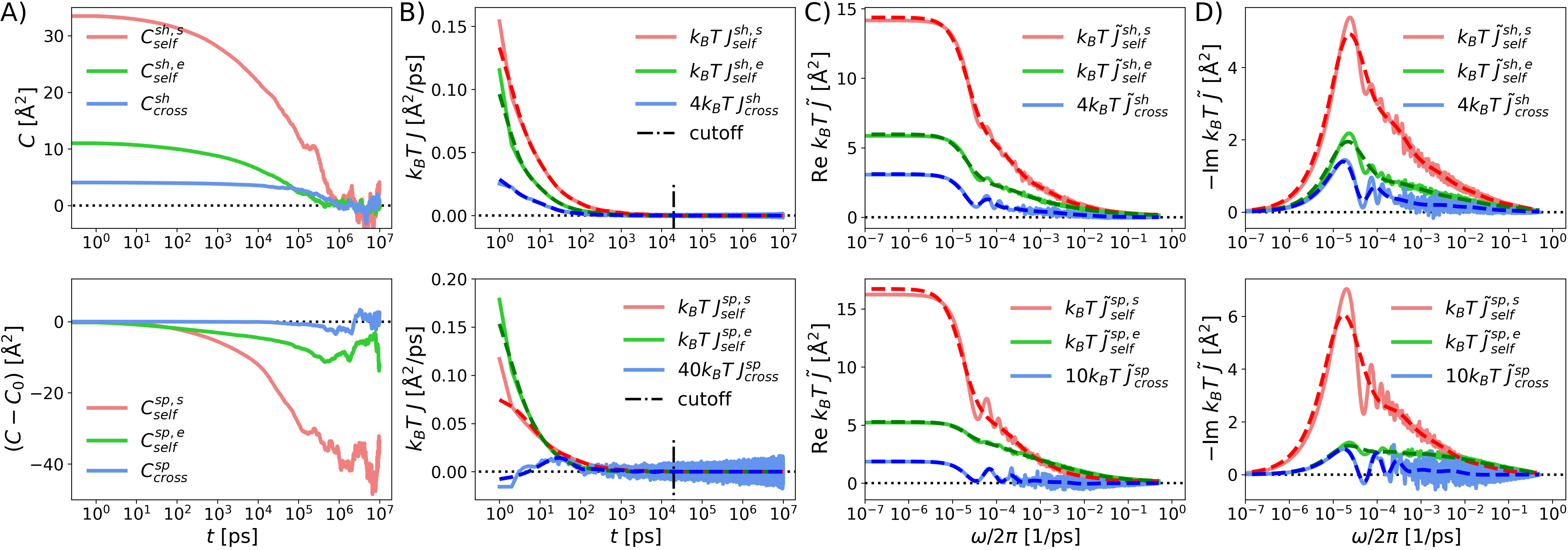}
\caption{\label{Fig-S3} Results for the  shift  (top) and splay modes (bottom)  of the wild-type coiled coil with capped neutral termini (ACE-CC-CONH$_2$).
 A) Self and cross-correlation functions $C_\mathrm{self}$ and $C_\mathrm{cross}$. For the splay mode, we subtract $C_\mathrm{0} = C(t=0)$ in order  to be able to compare all correlation functions in one plot. 
B)  Response functions $J(t)$, obtained from the numerical time-derivatives of the correlation functions, are shown as solid lines. 
The vertical dash-dotted lines represent the cutoff beyond which the response functions are set to zero to prevent noise artifacts when calculating Fourier transforms. 
C) Real and D) imaginary parts of $\Tilde{J}(\omega)$ obtained from discrete fast-Fourier transformation  of the time-domain  response function $J(t)$, 
 shown as solid lines. Dashed lines represent simultaneous  fits to the logarithmically-spaced real and imaginary parts of $\Tilde{J}(\omega)$  
 by a sum of 10 Debye  functions, for further details, see Section \ref{App3}. Dashed lines in panel B represent the inverse Fourier transform of  the fits in C and D.
 Note that the cross response functions in B, C, and D are multiplied with different scaling factors for a better visualization.} 
\end{figure*}
\clearpage
\begin{figure*}[]
\centering
\includegraphics[width=1.0\textwidth]{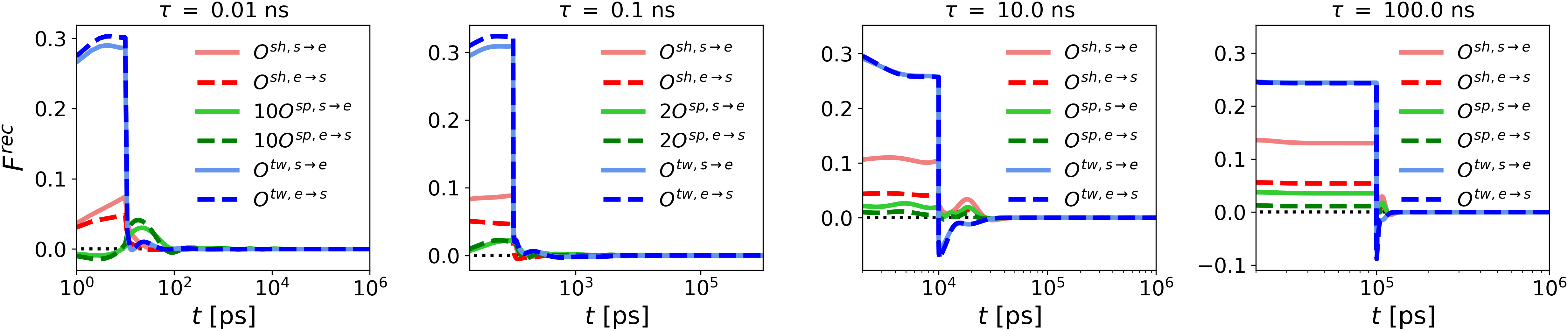}
\caption{\label{Fig-S5} Transmitted force profiles for  rectangular pulse force input for shift, splay, and twist modes of the wild-type coiled coil with capped neutral termini (ACE-CC-CONH$_2$) 
for different input pulse widths $\tau$ given on the top of each plot. Note that for the splay mode, transmitted force profiles are multiplied with a different scaling factor in each plot for a better visualization.}
\end{figure*}

\clearpage

%